\documentclass[12pt]{article}
\usepackage{amssymb}
\usepackage{graphicx}
\usepackage{amsthm}
\usepackage{psfrag}
\usepackage{dsfont}

\newtheoremstyle{custom}
{15pt}
{15pt}
{}
{}
{\bfseries}
{}
{.5em}
{}
\theoremstyle{custom}
\newtheorem{definition}{Definition}
\newtheorem{proposition}[definition]{Proposition}
\newtheorem{lemma}[definition]{Lemma}
\newtheorem{example}[definition]{Example}
  
\begin{document}

Published: Richard A Barry and Susan M Scott 2011 Class. Quantum Grav. 28(16) 165003 doi: 10.1088/0264-9381/28/16/165003
\\

\begin{center}\textbf{The Attached Point Topology of the Abstract Boundary For Space-Time}
\\

Richard A Barry and Susan M Scott
\\

\textit{Centre for Gravitational Physics, College of Physical and Mathematical Sciences, The Australian National University, Canberra ACT 0200, Australia}
\\

richard.barry@anu.edu.au, Susan.Scott@anu.edu.au\end{center}

\begin{abstract}Singularities play an important role in General Relativity and have been shown to be an inherent feature of most physically reasonable space-times.  Despite this, there are many aspects of singularities that are not qualitatively or quantitatively understood.  The abstract boundary construction of Scott and Szekeres has proven to be a flexible tool with which to study the singular points of a manifold.  The abstract boundary construction provides a `boundary' for any $n$-dimensional, paracompact, connected, Hausdorff, $C^{\infty}$ manifold.  Singularities may then be defined as entities in this boundary - the abstract boundary.  In this paper a topology is defined, for the first time, for a manifold together with its abstract boundary.  This topology, referred to as the attached point topology, thereby provides us with a description of how the abstract boundary is related to the underlying manifold.  A number of interesting properties of the topology are considered, and in particular, it is demonstrated that the attached point topology is Hausdorff.\end{abstract}

AMS classification scheme numbers: 53C23, 57R40, 57R15, 83C75

\section{Introduction}

Since the inception of the theory of General Relativity, singularities have played an important role.  In many instances, they were assumed to be an artefact of an idealised level of symmetry.  The powerful singularity theorems of Penrose and Hawking \cite{Hawking Penrose}, however, demonstrated that any generic space-time with a 'reasonable' distribution of matter satisfying physically reasonable conditions would necessarily contain singularities. This implied that singularities are therefore an integral part of a space-time.

Despite this, without the aid of any additional mathematical structure, we cannot fully answer the question ``what is a singularity?''. In part, this is due to the fact that a singularity is not, technically, part of the manifold, and therefore any description of it purely in terms of the manifold itself will not be complete.  An amount of extra mathematical structure is required in order to properly describe a singularity. This extra detail is provided by a boundary construction which gives us a way of rigorously describing the singular points of a manifold. A boundary construction is therefore an essential tool in properly understanding the global structure of a space-time.

Previously, there have been numerous attempts to produce a boundary construction for space-times - most notably the $g$-boundary of Geroch \cite{Geroch}, the $b$-boundary of Schmidt \cite{Schmidt} and the $c$-boundary of Geroch, Kronheimer and Penrose \cite{Geroch1}.  All of these boundary constructions, however, suffer from problems and limitations in terms of their application and physical results and, as such, they do not fully encapsulate all aspects of a singularity.  For a detailed summary of these constructions, see \cite{Hawking}, \cite{Ashley} and \cite{Whale}. The abstract boundary (a-boundary) construction of Scott and Szekeres \cite{Scott} offers an alternative to these constructions that is free of many of these issues.  It should be noted that other boundary constructions have been presented recently.  Most notable among these constructions is the iso-causal boundary of Garc\'{i}a-Parrado and Senovilla \cite{Senovilla} which uses an ideology similar to the a-boundary.  In addition, the c-boundary continues to be studied and numerous attempts have been made to address its known issues.  For a summary of these alternative c-boundary constructions, see \cite{c-boundary} and \cite{Flores}.

When dealing with abstract spaces, there is typically no predefined notion of how `close' or `separated' two elements of the space are relative to each other.  A topology provides us with such a notion and is therefore beneficial in understanding the structure of these spaces.  Although the abstract boundary construction provides us with a collection of abstract boundary points, without a topology on it we lack any sense of `where' these points are with respect to the manifold in question.  Since the abstract boundary points represent singularities (among other things), it is of obvious physical importance to know where these points are with respect to a space-time, and thus a topology on the manifold together with its abstract boundary is highly desirable.

It should be noted at this point that the $b$, $c$ and $g$-boundary constructions do have their own topologies.  In each case, however, there are problems associated with the separation of neighbouring points. The $b$-boundary, for instance, has been shown to identify the initial and final singularities of the closed Friedmann cosmology \cite{Bosshard}. It has also been shown that the $b$-boundary of a family of space-times, which includes the Friedmann and Schwarzschild solutions, is non-Hausdorff \cite{Johnson}.  Non-Hausdorff $g$-boundary constructions also occur naturally for many space-times.  As constructed in \cite{GLW}, these example space-times possess boundary points which are not $T_1$-separated from manifold points. The singular points are therefore arbitrarily close to `interior' manifold points. The $c$-boundary likewise suffers from topological separation problems between manifold points and boundary points.  This lack of separation between points appears to be a non-physical property, as it is not clear if non-Hausdorff space-times are realistic \cite{hajicek}. It is therefore physically desirable for there to exist a natural Hausdorff topology for the abstract boundary construction.  For a more complete discussion of the various topological problems associated with each of these three boundary constructions, see \cite{Ashley}.

The main difficulty in constructing a topology for a manifold $\mathcal{M}$ and its abstract boundary $\mathcal{B}(\mathcal{M})$ is that the abstract boundary points are produced via embeddings of the manifold.  This means that the abstract boundary points exist in a space separate to the manifold $\mathcal{M}$.  A way of relating the abstract boundary points back to the manifold is therefore required if they are to be included in open sets that also include elements of $\mathcal{M}$.

As usual, there exist a number of possible topologies which can be put on $\mathcal{M}\cup\mathcal{B}(\mathcal{M})$, some of which will be Hausdorff and first countable.  Ideally, we desire a topology that is physically useful, i.e., the topology should be able to tell us, for example, `where' in $\mathcal{M}\cup\mathcal{B}(\mathcal{M})$ the singularities are located.  It is therefore essential that the chosen topology somehow relates elements of the abstract boundary back to $\mathcal{M}$.  The topology that is presented in section \ref{attached point section}, namely the attached point topology, was developed with this in mind.  This topology relies on the idea of an abstract boundary point being attached to an open set of $\mathcal{M}$, and it represents one of the more natural possible constructions. What it means for an abstract boundary point to be attached, and other related concepts, are discussed in section \ref{attached boundary sets}.  Various properties of the open and closed sets of the attached point topology are then discussed in sections \ref{open and closed sets section}, \ref{inclusion section} and \ref{properties section}.

Within this work, we use the following fact frequently and so formally present it here for ease of reference.  Let $g$ be a Riemannian metric on a manifold $\mathcal{M}$, and let $\Omega_{p,q}$ denote the set of piecewise smooth curves in $\mathcal{M}$ from $p$ to $q$.  For every curve $c\in\Omega_{p,q}$ with $c:[0,1]\rightarrow\mathcal{M}$ there is a finite partition $0=t_1<t_2<...<t_k=1$ such that $c\mid[t_i,t_{i+1}]$ is smooth for each $i$, $1\leq i\leq k-1$.  The Riemannian arc length of $c$ with respect to $g$ is then defined to be $L(c)=\sum_{i=1}^{k-1}\int_{t_i}^{t_{i+1}}\sqrt{g(c'(t),c'(t))}dt$, and the Riemannian distance function, $d(p,q)$, between $p$ and $q$ is then defined in terms of this by $d(p,q)=\textrm{inf}\{L(c):c\in\Omega_{p,q}\}\geq0$.  The most useful property of this distance function is that the open balls defined by $B_\epsilon(p)=\{q\in\mathcal{M}:d(p,q)<\epsilon\}$ form a basis for the manifold topology, and thus the topology induced by the Riemannian metric agrees with the manifold topology \cite{Lorentzian}.

\section{The Abstract Boundary}\label{abstract boundary}

The a-boundary will now be defined.  For a more complete discussion of the a-boundary, see \cite{Scott}.  It will be assumed that all manifolds used in the following work will be n-dimensional, paracompact, connected, Hausdorff and smooth (i.e., $C^{\infty}$).  The manifold topology will be employed throughout the paper unless explicitly stated otherwise.  The principle feature of the a-boundary construction is that of an envelopment.

\begin{definition}[Embedding]The function $\mathcal{\phi: M\rightarrow\widehat{M}}$ is an \textit{embedding}  if $\phi$ is a homeomorphism between $\mathcal{M}$ and $\phi(\mathcal{M})$, where $\phi(\mathcal{M})$ has the subspace topology inherited from $\widehat{\mathcal{M}}$.\end{definition}

\begin{definition}[Envelopment]An \textit{enveloped manifold} is a triple ($\mathcal{M,\widehat{M},\phi}$) where $\mathcal{M}$ and $\mathcal{\widehat{M}}$ are differentiable manifolds of the same dimension $n$ and $\phi$ is a $\mathcal{C^{\infty}}$ embedding $\mathcal{\phi: M\rightarrow\widehat{M}}$.  The enveloped manifold will also be referred to as an \textit{envelopment of $\mathcal{M}$ by $\mathcal{\widehat{M}}$}, and $\mathcal{\widehat{M}}$ will be called the \textit{enveloping manifold.}\end{definition}

\begin{definition}[Boundary point]A \textit{boundary point p} of an envelopment ($\mathcal{M,\widehat{M},\phi}$) is a point in the topological boundary of $\phi(\mathcal{M})$ in $\mathcal{\widehat{M}}$.  The set of all such p is thus given by $\partial(\phi(\mathcal{M}))=\overline{\phi(\mathcal{M})}\backslash\phi(\mathcal{M})$ where $\overline{\phi(\mathcal{M})}$ is the closure of $\phi(\mathcal{M})$ in $\mathcal{\widehat{M}}$.  The boundary points are then simply the limit points of the set $\phi(\mathcal{M})$ in $\widehat{\mathcal{M}}$ which do not lie in $\phi(\mathcal{M})$ itself.

The characteristic feature of a boundary point is that every open neighbourhood of it (in $\widehat{\mathcal{M}}$) has non-empty intersection with $\phi(\mathcal{M})$.\end{definition}

\begin{definition}[Boundary set]A \textit{boundary set B} is a non-empty set of such boundary points for a given envelopment, i.e., a non-empty subset of $\partial(\phi(\mathcal{M}))$.\end{definition}

It is important to note that different boundary points will arise with different envelopments of $\mathcal{M}$.  In order to continue, a notion of equivalence between boundary sets of different envelopments is required.  This equivalence is defined in terms of a covering relation.

\begin{figure}[htb!]
\centering%
\psfrag{manifold}{$\scriptstyle\widehat{\mathcal{M}}$}
\psfrag{e}{$\scriptstyle\phi(\mathcal{M})$}
\psfrag{f}{$\scriptstyle\phi'(\mathcal{M})$}
\psfrag{m}{$\scriptstyle\widehat{\mathcal{M}}$}
\psfrag{a}{$\scriptstyle\widehat{\mathcal{M}}'$}
\psfrag{g}{$\scriptstyle \phi \circ \phi'^{-1}(\mathcal{U'}\cap\phi'(\mathcal{M}))$}
\includegraphics{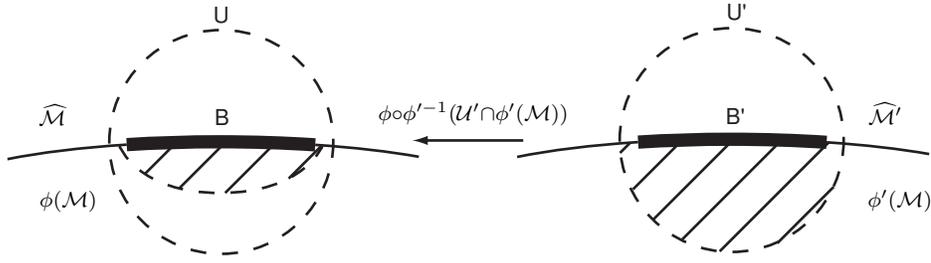}
\caption{the boundary set $B$ covers the boundary set $B'$}
\label{covering relation}
\end{figure}

\begin{definition}[Covering relation]Given a boundary set $B$ of one envelopment ($\mathcal{M,\widehat{M},\phi}$) and a boundary set $B'$ of a second envelopment ($\mathcal{M,\widehat{M'},\phi'}$), then \textit{$B$ covers $B'$} if for every open neighbourhood $\mathcal{U}$ of $B$ in $\mathcal{\widehat{M}}$ there exists an open neighbourhood $\mathcal{U'}$ of $B'$ in $\mathcal{\widehat{M'}}$ such that \begin{displaymath} \phi \circ \phi'^{-1}(\mathcal{U'}\cap\phi'(\mathcal{M}))\subset\mathcal{U}. \end{displaymath}
In essence, this definition says that a sequence of points from within $\mathcal{M}$ cannot get close to points of $B'$ without at the same time getting close to points of $B$.  See Fig \ref{covering relation}.\end{definition}

\begin{definition}[Equivalent]The boundary sets $B$ and $B'$ are \textit{equivalent} (written $B\sim B'$) if $B$ covers $B'$ and $B'$ covers $B$.  This definition produces an equivalence relation on the set of all boundary sets.  An equivalence class is denoted by $[B]$, where $B$ is a representative of the set of equivalent boundary sets under the covering relation.\end{definition}

\begin{definition}[Abstract boundary point]An \textit{abstract boundary point} is then defined to be an equivalence class $[B]$ that has a singleton point ${p}$ as a representative member.  Such an equivalence class will then be denoted by $[p]$. The set of all such abstract boundary points of a manifold $\mathcal{M}$ will be denoted by $\mathcal{B(M)}$ and called the \textit{abstract boundary} of $\mathcal{M}$.  The union of all points of a manifold $\mathcal{M}$ and its collection of abstract boundary points $\mathcal{B(M)}$ will be labelled as $\overline{\mathcal{M}}$, i.e., $\overline{\mathcal{M}}=\mathcal{M}\cup\mathcal{B}(\mathcal{M})$.\end{definition}

\section{Attached Boundary Points and Sets}\label{attached boundary sets}

In this section, a number of definitions will be presented that describe how the abstract boundary points of a manifold, $\mathcal{M}$, may be topologically related to the points of $\mathcal{M}$.

\begin{figure}[htb!]
\centering%
\psfrag{manifold}{$\widehat{\mathcal{M}}$}
\psfrag{e}{$\phi(\mathcal{M})$}
\psfrag{V}{$\phi(U)$}
\psfrag{U}{$\phi(V)$}
\includegraphics{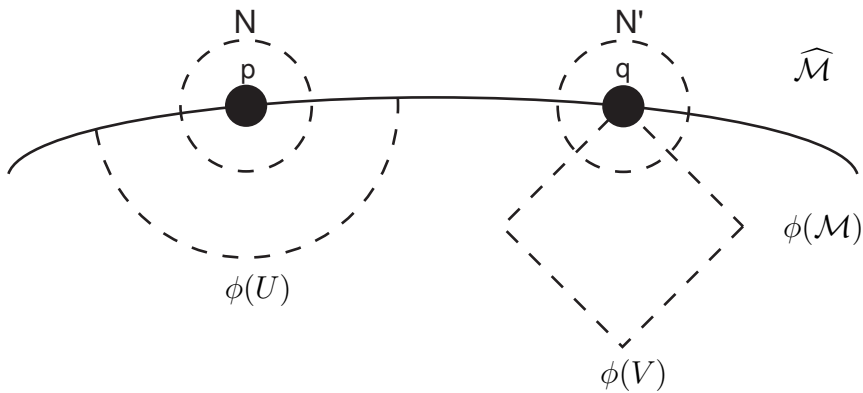}
\caption{boundary points $p$ and $q$ are attached to the open sets $U$ and $V$ respectively}
\label{attached}
\end{figure}

\begin{definition}[Attached boundary point]\label{attached point}Given an open set $U$ of $\mathcal{M}$ and an envelopment $\phi:\mathcal{M}\rightarrow \widehat{\mathcal{M}}$, then a boundary point $p$ of $\partial (\phi(\mathcal{M}))$ is said to be \textit{attached to} $U$ if every open neighbourhood $N$ of $p$ in $\widehat{\mathcal{M}}$ has non-empty intersection with $\phi(U)$, i.e., $N\cap\phi(U)\neq\emptyset$.  See Fig \ref{attached}.\end{definition}

\begin{figure}[htb!]
\centering%
\psfrag{manifold}{$\widehat{\mathcal{M}}$}
\psfrag{e}{$\phi(\mathcal{M})$}
\psfrag{V}{$\phi(U)$}
\psfrag{U}{$\phi(V)$}
\includegraphics{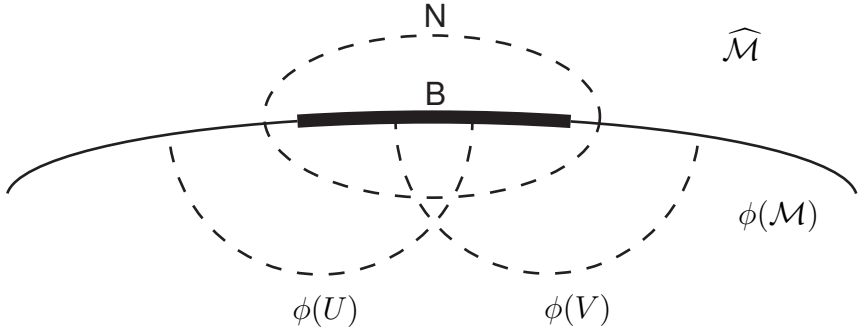}
\caption{boundary set $B$ is attached to the open sets $U$ and $V$}
\label{set}
\end{figure}

\begin{definition}[Attached boundary set]\label{attached set} Given an open set $U$ of $\mathcal{M}$ and an envelopment $\phi:\mathcal{M}\rightarrow \widehat{\mathcal{M}}$, then a boundary set $B\subset\partial (\phi(\mathcal{M}))$ is said to be \textit{attached to} $U$ if every open neighbourhood $N$ of $B$ in $\widehat{\mathcal{M}}$ has non-empty intersection with $\phi(U)$, i.e., $N\cap\phi(U)\neq\emptyset$.  See Fig \ref{set}.  Note that this does not necessarily imply that all points $q\in B$ are attached to $U$, as can be seen in the case illustrated by Fig \ref{set}. It does ensure, however, that at least one boundary point $p$ in $B$ is attached to $U$.\end{definition}

\begin{lemma}If $B\subset\partial(\phi(\mathcal{M}))$ is attached to an open set $U$ of $\mathcal{M}$, then there exists a $p\in B$ such that $p$ is attached to $U$.\end{lemma}

\textit{Proof:} The boundary set $B$ is attached to $U$.  Therefore, for every open neighbourhood $N$ of $B$ we have that $N\cap \phi(U)\neq\emptyset$.  Now assume that no point $q\in B$ is attached to $U$.  There therefore exists, for each $q$, an open neighbourhood $N_q$ of $q$ such that $N_q\cap \phi(U)=\emptyset$.  Now take the union $\bigcup_{q\in B} N_q$ of all of the $N_q$.  This is an open set containing $B$ such that $(\bigcup_{q\in B} N_q) \cap \phi(U)=\emptyset$.  This contradicts the fact that $B$ is attached to $U$, and therefore we have that some $q \in B$ must be attached to $U$. $\Box$
\\

Because boundary points which are equivalent may appear in a number of different envelopments, it is necessary to check that definitions (\ref{attached point}) and (\ref{attached set}) are well defined under the equivalence relation.  More specifically, we wish to show that if a boundary set $B\subset\partial(\phi(\mathcal{M}))$ is attached to an open set $U\subset\mathcal{M}$ and there exists a boundary set $B'\subset\partial(\psi(\mathcal{M}))$ that is equivalent to $B$, then $B'$ is also attached to $U$.

\begin{proposition}\label{B attached} Let $B\subset\partial(\phi(\mathcal{M}))$ be attached to an open set $U\subset\mathcal{M}$, and let $B'\subset\partial(\phi'(\mathcal{M}))$ be a boundary set of a second envelopment $\phi'$.  If $B'$ covers $B$, then $B'$ is also attached to the open set $U\subset\mathcal{M}$.\end{proposition}

\textit{Proof:} Let $B\subset\partial(\phi(\mathcal{M}))$ be attached to an open set $U\subset\mathcal{M}$, and let $B'\subset\partial(\phi'(\mathcal{M}))$ be a boundary set which covers $B$.  Assume that $B'$ is not attached to $U$.  Thus there exists an open neighbourhood $N$ of $B'$ in $\widehat{\mathcal{M}}'$ such that $N\cap\phi'(U)=\emptyset$.  Since $B'$ covers $B$, for every open neighbourhood $N'$ of $B'$ there exists an open neighbourhood $D$ of $B$ such that $\phi'\circ\phi^{-1}(D\cap\phi(\mathcal{M}))\subset N'$. This definition must be true for any neighbourhood $N'$ of $B'$, and so we choose $N'$ to be $N$, so that $\phi'\circ\phi^{-1}(D\cap\phi(\mathcal{M}))\subset N$.  Since $B$ is attached to $U$, $D\cap\phi(U)\neq\emptyset$, and since $D\cap\phi(U)\subset D\cap\phi(\mathcal{M})$, $\phi'\circ\phi^{-1}(D\cap\phi(U))\subset N$.  Now $D\cap\phi(U)\subset\phi(U)$ so that $\phi'\circ\phi^{-1}(D\cap\phi(U))\subset\phi'(U)$.  Since $\phi'\circ\phi^{-1}(D\cap\phi(U))\neq\emptyset$, it follows that $N\cap\phi'(U)\neq\emptyset$.  A contradiction is thus obtained as it was originally assumed that $N\cap\phi'(U)=\emptyset$.  $\Box$

\begin{definition}[Attached abstract boundary point]\label{attached abstract} The abstract boundary point $[p]$ is \textit{attached to} the open set $U$ of $\mathcal{M}$ if the boundary point $p$ is attached to $U$.\end{definition}

\textbf{Remark:} The abstract boundary point $[p]$ is an equivalence class of boundary sets which are equivalent to $p$.  By proposition (\ref{B attached}) the attached abstract boundary point definition is well defined as any boundary set $B$ such that $B\sim p$ is also attached to $U$, i.e., all members of the equivalence class $[p]$ are attached to $U$.

\begin{proposition}Given an open set $U$ of $\mathcal{M}$ and an envelopment $\phi:\mathcal{M}\rightarrow\widehat{\mathcal{M}}$, then the set $B_U$ of boundary points of $\partial(\phi(\mathcal{M}))$ which are attached to $U$ is closed in the induced topology on $\partial(\phi(\mathcal{M}))$. \end{proposition}

\textit{Proof:}  If $B_U=\emptyset$ or $\partial(\phi(\mathcal{M}))$ then, clearly, it is closed in the induced topology on $\partial(\phi(\mathcal{M}))$. So we will assume that $B_U\neq\emptyset$ or $\partial(\phi(\mathcal{M}))$.  For $B_U$ to be closed in the induced topology on $\partial(\phi(\mathcal{M}))$, then $\partial(\phi(\mathcal{M}))\backslash B_U\neq\emptyset$ must be open in $\partial(\phi(\mathcal{M}))$.  $\partial(\phi(\mathcal{M}))\backslash B_U$ contains the points $q\in\partial(\phi(\mathcal{M}))$ that are not attached to $U$, and thus there exists an open neighbourhood $N_q$ in $\widehat{\mathcal{M}}$ for each $q$ such that $N_q\cap\phi(U)=\emptyset$.  It follows that $N_q\cap B_U=\emptyset$ for each $q$, because otherwise $N_q$ would be a neighbourhood for some $p\in B_U$ and would thus intersect $\phi(U)$.  Call the union of all such $N_q$ neighbourhoods, $A$.  We therefore have that $A\cap\partial(\phi(\mathcal{M}))=\partial(\phi(\mathcal{M}))\backslash B_U$ is an open set in the induced topology on $\partial(\phi(\mathcal{M}))$ and therefore that $B_U$ is closed in $\partial(\phi(\mathcal{M}))$.  $\Box$

\begin{figure}[htb!]
\centering%
\psfrag{manifold}{$\widehat{\mathcal{M}}$}
\psfrag{e}{$\phi(\mathcal{M})$}
\psfrag{B}{$B_U$ (closed)}
\psfrag{U}{$\phi(U)$}
\includegraphics{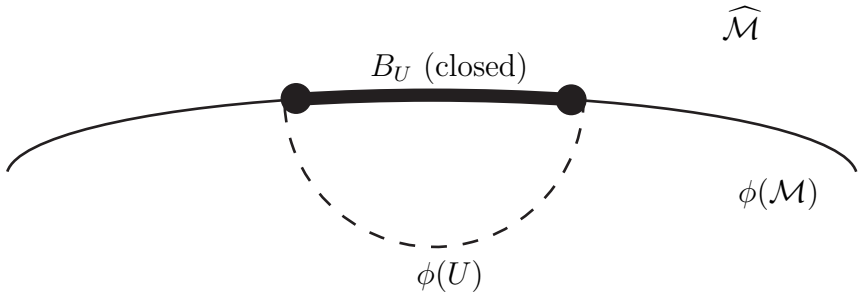}
\caption{the closed boundary set $B_U$ attached to $U$}
\label{B closed}
\end{figure}

\begin{proposition}Given an open set $U$ of $\mathcal{M}$ and an envelopment $\phi:\mathcal{M}\rightarrow\widehat{\mathcal{M}}$, then the set $B_U$ of boundary points of $\partial(\phi(\mathcal{M}))$ which are attached to $U$ is closed in $\widehat{\mathcal{M}}$. See Fig \ref{B closed}.  \end{proposition}

\textit{Proof:}  Once again, if $B_U=\emptyset$ or $\partial(\phi(\mathcal{M}))=\widehat{\mathcal{M}}\backslash(\widehat{\mathcal{M}}\backslash\overline{\phi(\mathcal{M})}\cup\phi(\mathcal{M}))$ then it is closed in $\widehat{\mathcal{M}}$, and so we will assume that $B_U\neq\emptyset$ or $\partial(\phi(\mathcal{M}))$.  If $\overline{B_U}=B_U$, then $B_U$ is closed in $\widehat{\mathcal{M}}$.  Let $x\in\widehat{\mathcal{M}}\backslash B_U$ and assume that $x$ is a limit point of $B_U$.  Since $\phi(\mathcal{M})$ and $\widehat{\mathcal{M}}\backslash\overline{\phi(\mathcal{M})}$ are open sets in $\widehat{\mathcal{M}}$, it is clear that $x\not\in\phi(\mathcal{M})$ and $x\not\in\widehat{\mathcal{M}}\backslash\overline{\phi(\mathcal{M})}$ else otherwise there would exist an open neighbourhood of $x$ which does not intersect $\partial(\phi(\mathcal{M}))$, and thus does not intersect $B_U$.  It follows that $x\in\partial(\phi(\mathcal{M}))$.  Since $x$ is a limit point of $B_U$, $N_x\cap B_U\neq\emptyset$ for every open neighbourhood $N_x$ of $x$, and therefore $N_x\cap\phi(U)\neq\emptyset$ for every $N_x$, because every $p\in B_U$ is attached to $U$.  This implies that $x$ is attached to $U$ which is a contradiction since it was originally assumed that $x\in\widehat{\mathcal{M}}\backslash B_U$.  It therefore follows that $B_U=\overline{B_U}$ and thus $B_U$ is closed in $\widehat{\mathcal{M}}$.  $\Box$

\section{The Attached Point Topology}\label{attached point section}

A topology on $\overline{\mathcal{M}}=\mathcal{M}\cup\mathcal{B}(\mathcal{M})$ may be constructed by defining the open sets in terms of the attached abstract boundary point definition (definition (\ref{attached abstract})). In keeping with the notion of constructing a natural topology, the open sets of $\mathcal{M}$ to which the abstract boundary points are attached are therefore taken to be the open sets of the manifold topology.

Consider the sets $A_i=U_i\cup B_i$, where $U_i$ is a non-empty open set of the manifold topology in $\mathcal{M}$ and $B_i$ is the set of all abstract boundary points which are attached to $U_i$. $B_i$ may be the empty set if no abstract boundary points are attached to $U_i$.  Consider also the sets $C_i$, where each $C_i$ is some subset of the abstract boundary $\mathcal{B}(\mathcal{M})$.  The collection of every $C_i$ set is the set of all subsets of the abstract boundary $\mathcal{B}(\mathcal{M})$, including all singleton sets $\{[p]\}$ where $[p]\in\mathcal{B}(\mathcal{M})$.  It will be seen (proposition \ref{attached boundary topology}) that the open sets of the topology induced on $\mathcal{B}(\mathcal{M})$ from the attached point topology on $\overline{\mathcal{M}}$ are precisely the $C_i$ sets.  Furthermore, it is the presence of certain $C_i$ sets which will ensure that the attached point topology is Hausdorff.

Let $\mathcal{V}$ be the set comprised of every $A_i$ set and every $C_i$ set. That is,

\begin{displaymath}
\mathcal{V}= \left\{ \begin{array}{ll}
A_i=U_i\cup B_i\\
C_i\subseteq\mathcal{B}(\mathcal{M})\\ \end{array} \right\}.
\end{displaymath}

\begin{lemma}\label{chart attached}Every abstract boundary point $[p]$ is attached to an open set $U_i$.\end{lemma}

\textit{Proof:} Let $N$ be any open neighbourhood of $p\in\partial(\phi(\mathcal{M}))$ in $\widehat{\mathcal{M}}$.  Since $p$ is a boundary point, every open neighbourhood of it has non-empty intersection with $\phi(\mathcal{M})$, and hence $N\cap\phi(\mathcal{M})$ is a non-empty open set in the subspace topology on $\phi(\mathcal{M})$.  In addition, since $\phi$ is an embedding, the non-empty set $U_i=\phi^{-1}(N\cap\phi(\mathcal{M}))$ is open in $\mathcal{M}$.  Now take any other open neighbourhood $N'$ of $p$ in $\widehat{\mathcal{M}}$.  Such a neighbourhood will always have non-empty intersection with $N\cap\phi(\mathcal{M})$.  This follows from the fact that the intersection of two open sets is another open set: $N'$ is an open set that contains $p$, and thus $N\cap N'=N^*$ is an open set that also contains $p$.  Because $N^*$ is a neighbourhood of $p$ we have that $N^*\cap\phi(\mathcal{M})\neq\emptyset$.  This then implies that $(N\cap N')\cap\phi(\mathcal{M})\neq\emptyset$, i.e., $N'\cap\phi(U_i)\neq\emptyset$.  This then is a statement of the attached boundary point condition, i.e., $p$ and thus $[p]$ is attached to $U_i$.  Every $[p]$ is therefore attached to an open set $U_i$ in $\mathcal{M}$. $\Box$

\begin{proposition}\label{1} The elements of $\mathcal{V}$ form a basis for a topology on $\overline{\mathcal{M}}$.\end{proposition}

\textit{Proof:} By definition, $\mathcal{M}$ is covered by the collection $\{U_i\}$ of open sets in $\mathcal{M}$.  Also, by lemma (\ref{chart attached}), each abstract boundary point is attached to an open set.  The set of open sets in $\mathcal{M}$ and their attached abstract boundary points, i.e., $\{A_i\}$, therefore covers $\overline{\mathcal{M}}$.

\begin{figure}[htb!]
\centering%
\psfrag{m}{$\widehat{\mathcal{M}}$}
\psfrag{e}{$\phi(\mathcal{M})$}
\psfrag{a}{$\phi(U_1)$}
\psfrag{b}{$\phi(U_2)$}
\includegraphics{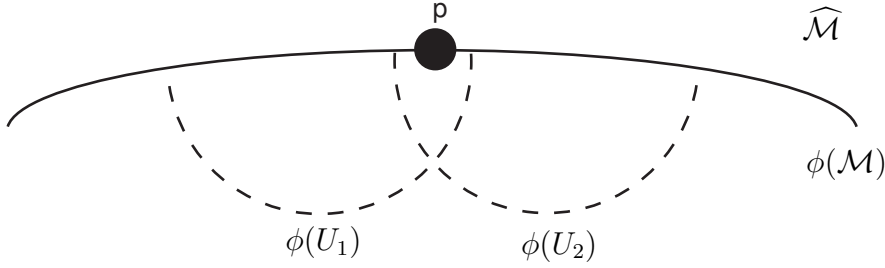}
\caption{subcase iia)}
\label{situation a}
\end{figure}

\begin{figure}[htb!]
\centering%
\psfrag{manifold}{$\widehat{\mathcal{M}}$}
\psfrag{e}{$\phi(\mathcal{M})$}
\psfrag{a}{$\phi(U_1)$}
\psfrag{b}{$\phi(U_2)$}
\includegraphics{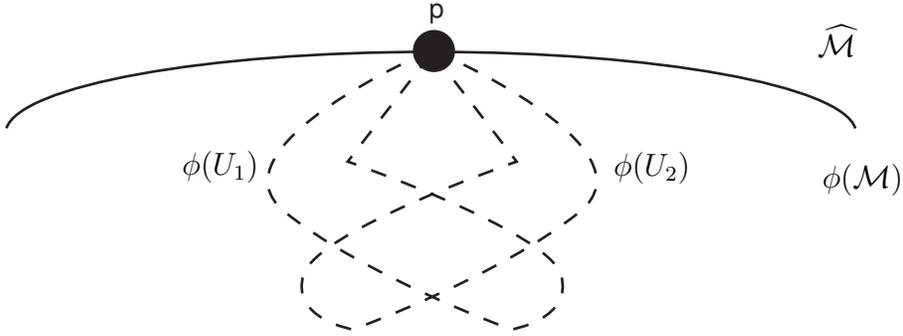}
\caption{subcase iib)}
\label{situation b}
\end{figure}

Now the intersection between two elements of $\mathcal{V}$ must be examined.  In doing so, there are three types of intersection that need to be considered.  The first is the intersection between $A_1=U_1\cup B_1$ and $A_2=U_2\cup B_2$. For this particular intersection, there are several cases to check:

\begin{enumerate}\item $U_1\cap U_2 \neq\emptyset$, $B_1\cap B_2=\emptyset$ (this includes the cases when $B_1=\emptyset$ or $B_2=\emptyset$)
\item $U_1\cap U_2\neq\emptyset$, $B_1\cap B_2\neq\emptyset$
\item $U_1\cap U_2=\emptyset$, $B_1\cap B_2\neq\emptyset$\end{enumerate}

i) In the first case we have that $U_1\cap U_2\neq\emptyset$ and $B_1\cap B_2=\emptyset$, and therefore $A_1\cap A_2=(U_1\cap U_2)\cup(B_1\cap B_2)$.  $U_1\cap U_2$ is another open set $U_3$. Assume there exists an abstract boundary point $[p]$ that is attached to $U_3$.  $[p]$ is therefore attached to $U_1$ ($[p]\in B_1$) and $U_2$ ($[p]\in B_2$) which would imply that $B_1\cap B_2\neq\emptyset$.  It thus follows that no abstract boundary point is attached to $U_3$ and so $A_1\cap A_2=U_3\cup B_3\in\mathcal{V}$ (where $B_3=\emptyset$).
\\

ii) There are two subcases that need to be considered in the case that $U_1\cap U_2\neq\emptyset$ and $B_1\cap B_2\neq\emptyset$. The first situation, subcase iia), is depicted in Fig \ref{situation a}, and the second situation, subcase iib), is depicted in Fig \ref{situation b}.

Subcase iia) refers to the situation where every abstract boundary point $[p]\in B_1\cap B_2$ is attached to $U_1\cap U_2$, and subcase iib) refers to the situation where $B_1\cap B_2\neq\emptyset$ and there exists a $[p]\in B_1\cap B_2$ which is not attached to $U_1\cap U_2$.

Let $I=A_1\cap A_2=(U_1\cap U_2)\cup(B_1\cap B_2)$, and $Q=(U_1\cap U_2)\cup B_{(U_1\cap U_2)}$ where $B_{(U_1\cap U_2)}$ is the set of all abstract boundary points which are attached to $U_1\cap U_2$. It may be the case that $B_{(U_1\cap U_2)}=\emptyset$ (subcase iib)). Otherwise, let $[p]$ be an abstract boundary point that is attached to $U_1\cap U_2$.  $[p]$ is therefore attached to $U_1$ ($[p]\in B_1$) and $U_2$ ($[p]\in B_2$).  We thus have that $[p]\in B_1\cap B_2$ and so $B_{(U_1\cap U_2)}\subseteq B_1\cap B_2$, in which case $Q\subseteq I$.  Since $Q\in\mathcal{V}$, any $x\in Q$ is contained in an element of $\mathcal{V}$ which is a subset of $I$.  Now suppose that $x\in I\backslash Q$, i.e., $x$ is an abstract boundary point $[p]$ which is not attached to $U_1\cap U_2$ (subcase iib)).  The abstract boundary point $[p]$ forms a set $C_i=\{[p]\}\in\mathcal{V}$, i.e., $x\in C_i\subseteq I$.  We therefore have that all elements of $I$ are contained in elements of $\mathcal{V}$, which are subsets of $I$.
\\

\begin{figure}[htb!]
\centering%
\psfrag{m}{$\widehat{\mathcal{M}}$}
\psfrag{e}{$\phi(\mathcal{M})$}
\psfrag{a}{$\phi(U_1)$}
\psfrag{b}{$\phi(U_2)$}
\includegraphics{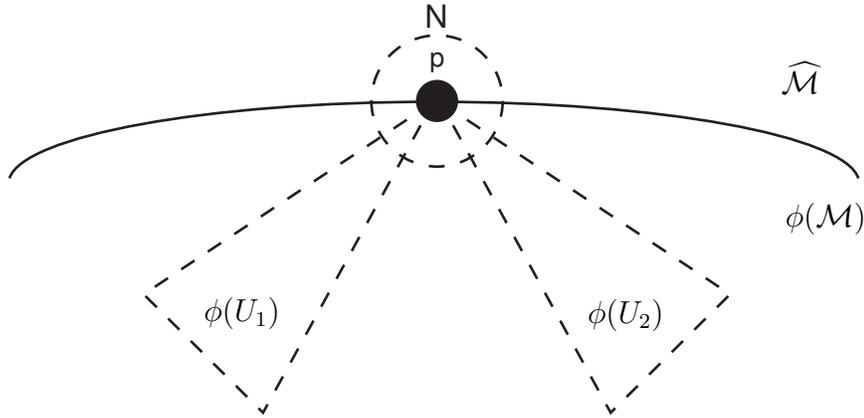}
\caption{case iii)}
\label{no intersection}
\end{figure}

$iii)$ Now consider the final case where $U_1\cap U_2=\emptyset$ and $B_1\cap B_2\neq\emptyset$.  See Fig \ref{no intersection}.  In this case we have that $A_1\cap A_2=B_1\cap B_2$, i.e., the intersection is a collection of abstract boundary points.  Since the $C_i$ sets are subsets of $\mathcal{B}(\mathcal{M})$, this collection of abstract boundary points will correspond to a $C_i$ set.

The next type of intersection to consider is the intersection between $A_i=U_i\cup B_i$ and $C_j$.  If $A_i\cap C_j\neq\emptyset$, then it consists of a collection of abstract boundary points.  As before, this collection of abstract boundary points will coincide with one of the $C_k$ sets and we will have that $(U_i\cup B_i)\cap C_j=C_k\in\mathcal{V}$.
\\

Finally, we consider the intersection between two $C_j$ sets.  Given any two $C_j$ sets, $C_i$ and $C_k$, that have non-empty intersection, then $C_i\cap C_k$ will be a set of abstract boundary points.  Since, by definition, the $C_j$ sets are subsets of $\mathcal{B}(\mathcal{M})$, there will always exist another set $C_l$ such that $(C_i\cap C_k)=C_l\in\mathcal{V}$.
\\

This concludes the proof that the elements of $\mathcal{V}$ form a basis for a topology on $\overline{\mathcal{M}}$. $\Box$

\begin{definition}[Attached point topology] \textit{The attached point topology} on $\overline{\mathcal{M}}$ is the topology on $\overline{\mathcal{M}}$ which has the basis $\mathcal{V}$.\end{definition}

The aim of the attached point topology is to investigate how a given abstract boundary point is related to the underlying manifold $\mathcal{M}$.  This is achieved by hardwiring into the topology, via the definition of an attached abstract boundary point (definition \ref{attached abstract}), what it means for an abstract boundary point to be `close' to some part of $\mathcal{M}$.  Basically, the location of a particular abstract boundary point $[p]$ is fully determined by the set of open sets of $\mathcal{M}$ to which it is attached.  This provides a natural motivation for our choice of topology on $\overline{\mathcal{M}}$ with the sets $A_i=U_i\cup B_i$, comprised of open sets $U_i$ of $\mathcal{M}$ together with all abstract boundary points which are attached to $U_i$, forming basis elements for the attached point topology on $\overline{\mathcal{M}}$.

It should also be noted that the $C_i$ sets are an important and necessary addition to the basis $\mathcal{V}$.  As was seen in case iii) in the proof of proposition \ref{1}, where $U_1\cap U_2=\emptyset$ and $B_1\cap B_2\neq\emptyset$ (see fig \ref{no intersection}), we have that $A_1\cap A_2=B_1\cap B_2$, i.e., the intersection is a collection of abstract boundary points.  So in any topology generated from a basis which includes the sets $A_i$, this collection of abstract boundary points is an open set.  This in turn forces the existence of basis elements which are collections of abstract boundary points, i.e., the $C_i$ sets.

\section{Open and Closed Sets in the Attached Point Topology}\label{open and closed sets section}

The open sets of $\overline{\mathcal{M}}$ consist of arbitrary unions of the elements of $\mathcal{V}$.  At first inspection it may seem that an arbitrary open set $(U_i\cup B_i)\cup(U_j\cup B_j)\cup...$ is another basis element $U_k\cup B_k$ because it is possible to write $(U_i\cup B_i)\cup(U_j\cup B_j)\cup...$ as $(U_i\cup U_j\cup...)\cup(B_i\cup B_j...)=U_k\cup(B_i\cup B_j...)$.  The following proposition demonstrates, however, that this is not true in general, as there may be abstract boundary points attached to $U_k$ that are not contained in $B_i\cup B_j...$, i.e., that are not attached to $U_i$, $U_j,...$ .

\begin{proposition}\label{M open and closed} The sets $\mathcal{M}$ and $\mathcal{B}(\mathcal{M})$ are each both open and closed in the attached point topology on $\overline{\mathcal{M}}$.\end{proposition}

\textit{Proof:} For a manifold $\mathcal{M}$, there exists a complete metric $d$ on $\mathcal{M}$ such that the topology induced by $d$ agrees with the manifold topology of $\mathcal{M}$ \cite{Lorentzian}.

Choose $\epsilon>0$, and for each $x\in\mathcal{M}$, let $U_x$ be the open ball $U_x=\{y\in\mathcal{M}:d(x,y)<\epsilon\}$. Now consider the envelopment $\phi:\mathcal{M}\rightarrow\widehat{\mathcal{M}}$ and a boundary point $p\in\partial(\phi(\mathcal{M}))$. We know that $p\notin\overline{\phi(U_x)}$ since $d$ is a complete metric on $\mathcal{M}$.  Thus the set $\widehat{\mathcal{M}}\backslash\overline{\phi(U_x)}$ is an open neighbourhood of $p$ in $\widehat{\mathcal{M}}$ which does not intersect $\phi(U_x)$, and so $p$ is not attached to $U_x$. It follows that no boundary point $p$ of any envelopment of $\mathcal{M}$ is attached to $U_x$, which implies that $U_x$ has no attached abstract boundary points, i.e., $B_x=\emptyset$.

Now \begin{eqnarray}\bigcup_{x\in\mathcal{M}} A_x&=&\bigcup_{x\in\mathcal{M}}(U_x\cup B_x)\nonumber\\
&=&(\bigcup_{x\in\mathcal{M}}U_x)\cup(\bigcup_{x\in\mathcal{M}}B_x)\nonumber\\
&=&\mathcal{M}\cup\emptyset=\mathcal{M}\nonumber.\end{eqnarray}

It follows that $\mathcal{M}$ is open in $\overline{\mathcal{M}}$ and thus $\mathcal{B}(\mathcal{M})$ is closed.  Since $\mathcal{B}(\mathcal{M})\subseteq \mathcal{B}(\mathcal{M})$, $\mathcal{B}(\mathcal{M})$ is a basis element $C_i$ and is therefore open in $\overline{\mathcal{M}}$, which means that $\mathcal{M}$ is closed.  So the sets $\mathcal{M}$ and $\mathcal{B}(\mathcal{M})$ are each both open and closed in the attached point topology on $\overline{\mathcal{M}}$.  $\Box$
\\

This proposition has demonstrated that $\mathcal{M}=\bigcup_{x\in\mathcal{M}}(U_x\cup B_x)$.  In general, $\mathcal{M}\neq\mathcal{M}\cup B_{\mathcal{M}}$, where $B_{\mathcal{M}}$ is the collection of all abstract boundary points attached to $\mathcal{M}$ (i.e., $B_{\mathcal{M}}=\mathcal{B}(\mathcal{M})$ as every boundary point of every envelopment of $\mathcal{M}$ is attached to $\mathcal{M}$).  It has therefore been demonstrated that an arbitrary union of basis elements of the topology is, in general, not another basis element.

\begin{figure}[htb!]
\centering%
\psfrag{p}{$p=(x_0,0)$}
\psfrag{m}{$\widehat{\mathcal{M}}$}
\psfrag{v}{$\phi(\mathcal{M})$}
\psfrag{x}{$x_1$}
\psfrag{b}{$U_1$}
\includegraphics{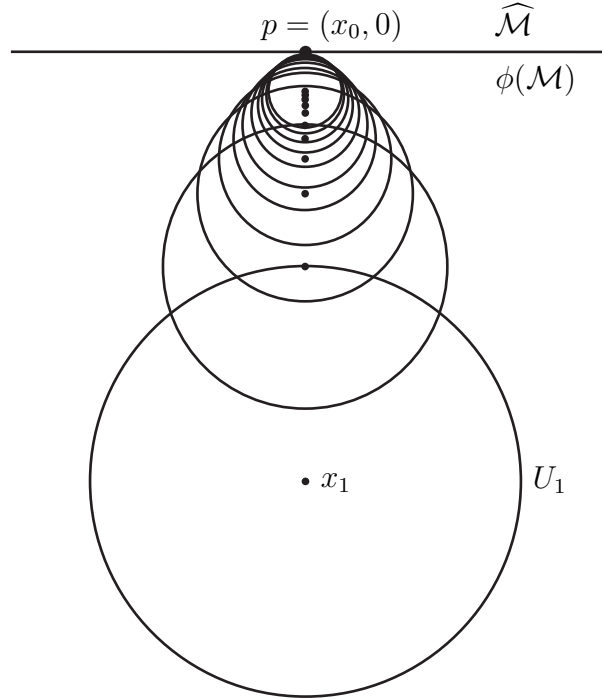}
\caption{the first $11$ elements of the sequence $\{x_n\}$ and their open ball neighbourhoods $U_n$}
\label{example}
\end{figure}

\begin{example}This example illustrates that $\mathcal{M}$ is not the only example of an open set in $\overline{\mathcal{M}}$ that has no attached abstract boundary points and may also be written as a union of $U_i\cup B_i$ basis sets.

Consider $\mathcal{M}=\{(x,y)\in\mathds{R}^2:y<0\}$, $\widehat{\mathcal{M}}=\mathds{R}^2$ and let $\phi:\mathcal{M}\rightarrow\widehat{\mathcal{M}}$ be the inclusion map.  Let $p$ be the boundary point $(x_0,0);p\in\partial(\phi(\mathcal{M}))$ is an abstract boundary point representative. Define a sequence $\{x_n\}$ of $\mathcal{M}$ by $x_n\equiv(x_0,-\frac{1}{n})$ so that $d(x_n,p)=1/n$, where $d$ is the distance function on $\widehat{\mathcal{M}}$ (which produces the manifold topology of $\mathds{R}^2$).  Around every point $x_n$ consider the open ball defined by $U_n=\{y\in\widehat{\mathcal{M}}:d(x_n,y)<1/(n+1)\}$ (see Fig \ref{example}).  By construction, for each $n$, $\overline{U_n}\subset\mathcal{M}$ and thus each $U_n$ has no attached abstract boundary points, i.e., $B_n=\emptyset$ and $U_n=U_n\cup B_n$. Because the sequence $\{x_n\}$ converges to the point $p$, it follows that every open neighbourhood of $p$ will contain some point $x_n$ and therefore will intersect the open ball $U_n$.  The abstract boundary point $[p]$ is therefore attached to $O=\bigcup_n U_n$, but $O$ may be expressed as a union of non-empty open sets $U_n$ in $\mathcal{M}$, each of which does not have any attached abstract boundary points, i.e., $O=\bigcup_n U_n=\bigcup_n(U_n\cup B_n)$.\end{example}

\begin{lemma}The singleton abstract boundary point sets, $\{[p]\}$, are both open and closed in the attached point topology on $\overline{\mathcal{M}}$.\end{lemma}

\textit{Proof:} For each abstract boundary point $[p]$, $\{[p]\}\subseteq \mathcal{B}(\mathcal{M})$.  Thus $\{[p]\}=C_i$, a basis element of $\mathcal{V}$, and is therefore open in the attached point topology on $\overline{\mathcal{M}}$.

Now $\mathcal{B}(\mathcal{M})\backslash\{[p]\}\subseteq \mathcal{B}(\mathcal{M})$ and is therefore a basis element $C_j$ of $\mathcal{V}$. By Proposition (\ref{M open and closed}), $\mathcal{M}$ is open in the attached point topology on $\overline{\mathcal{M}}$.  The set $\mathcal{M}\cup C_j=\mathcal{M}\cup (\mathcal{B}(\mathcal{M})\backslash\{[p]\})=\overline{\mathcal{M}}\backslash\{[p]\}$ is open in $\overline{\mathcal{M}}$ as it is the union of two open sets.  It follows that $\{[p]\}$ is closed in $\overline{\mathcal{M}}$.

Thus the singleton abstract boundary point sets, $\{[p]\}$, are both open and closed in the attached point topology on $\overline{\mathcal{M}}$.  $\Box$

\begin{proposition}\label{attached boundary topology} The open sets of the induced topology on $\mathcal{B}(\mathcal{M})\subset\overline{\mathcal{M}}$, where $\overline{\mathcal{M}}$ has the attached point topology, are the $C_i$ sets defined in the basis $\mathcal{V}$.\end{proposition}

\textit{Proof:} Let $T_{\overline{\mathcal{M}}}$ be the attached point topology on $\overline{\mathcal{M}}$.  The subspace topology on $\mathcal{B}(\mathcal{M})$ is the collection of sets $T_{\mathcal{B}(\mathcal{M})}=\{U\cap\mathcal{B}(\mathcal{M}):U\in T_{\overline{\mathcal{M}}}\}$. $T_{\overline{\mathcal{M}}}$ is the collection of arbitrary unions and finite intersections of $U_j\cup B_j$ and $C_i$ sets.  The intersection of these sets $U$ with $\mathcal{B}(\mathcal{M})$ is therefore the collection of $C_i$ sets.  $\Box$

\section{The Inclusion Map from $\mathcal{M}$ to $\overline{\mathcal{M}}$}\label{inclusion section}

We now consider the inclusion map $i:\mathcal{M}\rightarrow\overline{\mathcal{M}}=\mathcal{M}\cup \mathcal{B}(\mathcal{M})\mid i(p)=p$.  It can be shown that the inclusion map is an embedding.

\begin{proposition}\label{i embedding}  If $\overline{\mathcal{M}}$ has the attached point topology, then the inclusion mapping $i:\mathcal{M}\rightarrow\overline{\mathcal{M}}\mid i(p)=p$ is an embedding.\end{proposition}

\textit{Proof:} The inclusion mapping $i$ is an embedding if it is a homeomorphism of $\mathcal{M}$ onto $i(\mathcal{M})$ in the subspace topology on $i(\mathcal{M})\cap\overline{\mathcal{M}}$.  Clearly $i$ is a bijection of $\mathcal{M}$ onto $i(\mathcal{M})$.  Now let $T_\mathcal{M}$ be the usual topology on $\mathcal{M}$ consisting of the collection of open sets $\{U_i\}$, $T_{\overline{\mathcal{M}}}$ the attached point topology on $\overline{\mathcal{M}}$ as defined in section \ref{attached point section} from the basis elements of $\mathcal{V}$, i.e., $T_{\overline{\mathcal{M}}}$ is the collection of arbitrary unions and finite intersections of the $U_i\cup B_i$ and $C_j$ sets, and $T_{i(\mathcal{M})}$ the subspace topology on $i(\mathcal{M})\cap\overline{\mathcal{M}}$.  The subspace topology $T_{i(\mathcal{M})}$ is therefore the collection of sets $T_{i(\mathcal{M})}=\{U_k\}$.  Clearly both $i$ and $i^{-1}$ are continuous with respect to $T_{\mathcal{M}}$ and $T_{i(\mathcal{M})}$.  It has thus been demonstrated that $i:\mathcal{M}\rightarrow\overline{\mathcal{M}}\mid i(p)=p$ is a homeomorphism onto its image in the induced topology and thus it is an embedding. $\Box$
\\

Because it has been shown that $i:\mathcal{M}\rightarrow\overline{\mathcal{M}}\mid i(p)=p$ is an embedding, we may view $\overline{\mathcal{M}}$ as simply $\mathcal{M}$ with the addition of its abstract boundary points.  This is a pleasing result as one would expect the nature of $\mathcal{M}$ to be preserved in $\overline{\mathcal{M}}$.

The following properties of $i(\mathcal{M})$ are readily obtained.

\begin{lemma}\label{open inclusion}For the inclusion mapping $i:\mathcal{M}\rightarrow\overline{\mathcal{M}}\mid i(p)=p$, $i(\mathcal{M})$ is both open and closed in the attached point topology on $\overline{\mathcal{M}}$, $\overline{i(\mathcal{M})}\neq\overline{\mathcal{M}}$ and $\partial(i(\mathcal{M}))=\emptyset$.\end{lemma}

\textit{Proof:} Since $i(\mathcal{M})=\mathcal{M}$, it follows from proposition (\ref{M open and closed}) that $i(\mathcal{M})$ is both open and closed in the attached point topology on $\overline{\mathcal{M}}$.

Because $i(\mathcal{M})$ is closed, $\overline{i(\mathcal{M})}=i(\mathcal{M})=\mathcal{M}\neq\overline{\mathcal{M}}=\mathcal{M}\cup \mathcal{B}(\mathcal{M})$.  Now $\partial(i(\mathcal{M}))=\overline{i(\mathcal{M})}\backslash i(\mathcal{M})=i(\mathcal{M})\backslash i(\mathcal{M})=\emptyset$.  $\Box$
\\

In particular, lemma \ref{open inclusion} demonstrates that under the inclusion mapping $i$, $\mathcal{M}$ is open and thus $\mathcal{B}(\mathcal{M})$ is closed in the attached point topology on $\overline{\mathcal{M}}$ as one would desire.

\section{Properties of the Attached Point Topology}\label{properties section}

A number of important properties of the attached point topology will now be considered.

\begin{proposition}\label{attached hausdorff} The topological space $(\overline{\mathcal{M}}, T_{\overline{\mathcal{M}}})$, where $T_{\overline{\mathcal{M}}}$ is the attached point topology on $\overline{\mathcal{M}}$, is Hausdorff.\end{proposition}

\textit{Proof:} Consider two distinct points in $\mathcal{M}$, $x$ and $y$.  Because $\mathcal{M}$ is Hausdorff, there exist open neighbourhoods $N_x$ and $N_y$ of $x$ and $y$, respectively, such that $N_x\cap N_y=\emptyset$.  We now consider whether or not the topological space $(\overline{\mathcal{M}}, T_{\overline{\mathcal{M}}})$ is Hausdorff, for while the manifold $\mathcal{M}$ is defined to be Hausdorff, $(\overline{\mathcal{M}},T_{\overline{\mathcal{M}}})$ is not necessarily Hausdorff.

Given the existence of a complete metric $d$ on $\mathcal{M}$, it was demonstrated in the proof of proposition (\ref{M open and closed}) that, for any $v > 0$, the open ball $\{p\in\mathcal{M}:d(x,p)<v\}$, based at the point $x$ has no attached abstract boundary points.  Since $N_x$ is an open neighbourhood of $x$ in $\mathcal{M}$, it is possible to choose an $\epsilon>0$ such that for the open ball $U_x=\{p\in\mathcal{M}:d(x,p)<\epsilon\}$, $\overline{U_x}\subset N_x$.  Now the basis element of $\mathcal{V}$, $U_x\cup B_x$, is simply $U_x$ since $B_x=\emptyset$.

Likewise, we can choose an $\eta>0$ such that for the open ball $U_y=\{p\in\mathcal{M}:d(y,p)<\eta\}$, $\overline{U_y}\subset N_y$.  The basis element $U_y\cup B_y$ is simply $U_y$ since $B_y=\emptyset$.

Thus, $x\in U_x\cup B_x$, $y\in U_y\cup B_y$ and $(U_x\cup B_x)\cap(U_y\cup B_y)=(U_x\cap U_y)\subseteq N_x\cap N_y=\emptyset$.  The open sets $U_x\cup B_x$ and $U_y\cup B_y$ are therefore disjoint open neighbourhoods of $x$ and $y$ respectively.

Now consider a point $x\in\mathcal{M}$ and an abstract boundary point $[p]\in\mathcal{B}(\mathcal{M})$.  As before, for $\epsilon>0$, the open ball $U_x=\{p\in\mathcal{M}:d(x,p)<\epsilon\}$ based at the point $x$ has no attached abstract boundary points.  Thus, the basis element of $\mathcal{V}$, $U_x\cup B_x$ is simply $U_x$.  Now $C_i=\{[p]\}$ is also a basis element of $\mathcal{V}$, and $(U_x\cup B_x)\cap C_i=U_x\cap C_i=\emptyset$.  The open sets $U_x\cup B_x$ and $C_i$ are therefore disjoint open neighbourhoods of $x$ and $[p]$ respectively.

Finally, consider two distinct abstract boundary points $[p]$ and $[q]$, i.e., $p$ is not equivalent to $q$.  The basis elements of $\mathcal{V}$, $C_i=\{[p]\}$ and $C_j=\{[q]\}$, are disjoint open neighbourhoods of $[p]$ and $[q]$ respectively, since $[p]$ and $[q]$ are different equivalence classes.

Having considered all possible combinations of different types of elements of $\overline{\mathcal{M}}$, namely $x,y\in\mathcal{M}$, $x\in\mathcal{M}$ and $[p]\in\mathcal{B}(\mathcal{M})$, and $[p],[q]\in\mathcal{B}(\mathcal{M})$, we have thereby demonstrated that the topological space $(\overline{\mathcal{M}}, T_{\overline{\mathcal{M}}})$ is indeed Hausdorff. $\Box$
\\

We shall also check if the attached point topology is first countable.

\begin{proposition}The attached point topology on $\overline{\mathcal{M}}$ is first countable.\end{proposition}

\textit{Proof:} A topological space $X$ is said to be \textit{first countable} if for each $x\in X$, there exists a sequence $U_1$, $U_2$,... of open neighbourhoods of $x$ such that for any open neighbourhood, $V$, of $x$, there exists an integer, $i$, such that $U_i\subseteq V$.

For $X=\overline{\mathcal{M}}$, we firstly consider the case where $x\in\mathcal{M}$.  Given the existence of a complete metric $d$ on $\mathcal{M}$, we know from the proof of proposition (\ref{M open and closed}) that, for $n\in\mathds{N}$, the open balls $U_n=\{p\in\mathcal{M}:d(x,p)<1/n\}$ based at the point $x$ have no attached abstract boundary points.  The sets $U_n\cup B_n=U_n$ are basis elements of $\mathcal{V}$, and so $U_1$, $U_2$,... is a sequence of open neighbourhoods of $x$.

Let $V$ be an open neighbourhood of $x$ in $\overline{\mathcal{M}}$. $V$ is an arbitrary union or finite intersection of basis elements $A_i$ and $C_j$ and therefore has the form $V=U\cup B$ where $U$ is an open set in $\mathcal{M}$, $x\in U$, and $B\subseteq\mathcal{B}(\mathcal{M})$ (where possibly $B=\emptyset$). It is possible to choose an $n\in\mathds{N}$, such that, for the open ball $U_n$, $\overline{U_n}\subset U$.  Thus $U_n\subseteq V$. We have therefore shown that $\overline{\mathcal{M}}$ is first countable at $x$, for all $x\in\mathcal{M}$.

Now we consider an abstract boundary point $[p]\in\mathcal{B}(\mathcal{M})$.  For each $n\in\mathds{N}$, define $C_n=\{[p]\}$.  The basis elements $C_n$ form a sequence, $C_1$, $C_2$,... of open neighbourhoods of $[p]$.  Now if $V$ is an open neighbourhood of $[p]$ in $\overline{\mathcal{M}}$, then $[p]\in V$ and $C_n=\{[p]\}\subseteq V$.  This means that $\overline{\mathcal{M}}$ is first countable at $[p]$, for all $[p]\in\mathcal{B}(\mathcal{M})$.

We have thereby shown that the attached point topology for $\overline{\mathcal{M}}$ is first countable.  $\Box$

\section{Conclusion}

The abstract boundary construction is a mathematical tool used to find and classify the boundary features of a space-time, including any singularities.  The ability to classify singular points, however, represents only half of the picture.  In order to fully understand the significance of a particular singularity, we must also understand how that singularity is connected to the original space-time.  The attached point topology, defined on the union of a manifold with its abstract boundary, provides us with one such description, and has the advantage that its construction flows naturally from the definitions of the abstract boundary construction itself.

It was shown that the attached point topology is Hausdorff which is considered an important ingredient for a workable boundary definition.  One of the key elements in the attached point topology being Hausdorff is that every abstract boundary point is an open set.  As a consequence of this, every abstract boundary point may be separated from every other abstract boundary point as well as every point of the manifold $\mathcal{M}$. Therefore, as well as ensuring that $\mathcal{V}$ is a basis, the $C_i$ sets also serve to guarantee that the attached point topology is Hausdorff. The intention of the attached point topology was to construct a Hausdorff topology which flows naturally from the attached abstract boundary point definition (definition \ref{attached abstract}).  The defined $C_i$ sets represent a simple solution to the problem of defining a collection of sets of abstract boundary points which ensure that $\mathcal{V}$ is a basis for a Hausdorff topology.

The `location' of an abstract boundary point, e.g., a singularity, is hardwired into the attached point topology through the basis elements $A_i = U_i\cup B_i$. Every abstract boundary point is attached to a non-empty open set $U_i\subset\mathcal{M}$ (Lemma (\ref{chart attached})). This means that, for a given abstract boundary point, for a boundary point representative $p$ occurring in an envelopment $\phi$, the open set image $\phi(U_i)$ of $U_i$ under $\phi$ extends all the way out to $p$ in this envelopment. Thus the boundary point $p$ is `close' to the open set $U_i$.  Since this must also be true for every boundary point representative of the abstract boundary point, we thereby have an \textit{a priori} knowledge of which particular open sets of  $\mathcal{M}$ are `close' to our given abstract boundary point.  This gives us the \textit{location} for the boundary features such as singularities.

The fact that the attached point topology is naturally Hausdorff is a pleasing result as, unlike a number of the other boundary constructions, we do not have to be concerned with specific space-time examples where we lose separability, as was discussed in the introduction. In addition, we do not need to consider further conditions on the manifold itself or its boundary in order to ensure that the topology on $\overline{\mathcal{M}}$ is Hausdorff.  In the case of the $c$-boundary, for instance, it has been suggested that extra causality conditions on the manifold, such as it being stably causal, would ensure that the resulting topology on the boundary is Hausdorff \cite{Szabados}, \cite{Rube}.

In a forthcoming paper, a second topology will be considered for $\overline{\mathcal{M}}= \mathcal{M}\cup B(\mathcal{M})$ in which the abstract boundary $B(\mathcal{M})$ is a closed set. As a consequence of this, however, a number of the abstract boundary points become inseparable, and thus the Hausdorff property is lost in general.  Separability is lost in a very particular way, however, to the extent that this lack of separability may contain additional information about the abstract boundary itself.
\\

\textbf{Acknowledgements } 
\\

The authors would like to thank Benjamin Whale for useful discussions relating to this paper.


\end{document}